\documentclass[12pt]{article}

\usepackage{bbm}
\usepackage{graphicx}

\newcommand{\be}{\begin{equation}}
\newcommand{\ee}{\end{equation}}
\newcommand{\ba}{\begin{eqnarray}}
\newcommand{\ea}{\end{eqnarray}}
\newcommand{\no}{\nonumber\\}

\begin{document}

\title{Extension of the Buchalla--Safir bound}
\author{L.\ Lavoura\thanks{E-mail: balio@cfif.ist.utl.pt} \\
\small Universidade T\'ecnica de Lisboa \\
\small Centro de F\'\i sica das Interac\c c\~oes Fundamentais \\
\small Instituto Superior T\'ecnico, 1049-001 Lisboa, Portugal}

\date{18 February 2004}

\maketitle

\begin{abstract}
I provide a simple derivation of the Buchalla--Safir bound on $\gamma$.
I generalize it to the case where
an upper bound on the phase of the penguin pollution is assumed.
I apply the Buchalla--Safir bound,
and its generalization,
to the recent Belle data on $CP$ violation in $B \to \pi^+ \pi^-$.
\end{abstract}

\section{Introduction}

$CP$ violation in $B^0_d$--$\bar B^0_d$ mixing
and in the decays of those mesons to $\pi^+ \pi^-$
is parametrized by
\be
\lambda = \frac{q}{p}\, \frac{\bar A}{A},
\label{iurty}
\ee
where $q/p$ relates to $B^0_d$--$\bar B^0_d$ mixing,
$A$ is the amplitude for $B^0_d \to \pi^+ \pi^-$,
and $\bar A$ is the amplitude for $\bar B^0_d \to \pi^+ \pi^-$
\cite{livro}.
Two $CP$-violating quantities can be measured:
\ba
S &=& \frac{2 \mathrm{Im}\, \lambda}{1 + \left| \lambda \right|^2},
\\
C &=& \frac{1 - \left| \lambda \right|^2}{1 + \left| \lambda \right|^2}.
\ea

Let
\be
\frac{q}{p} = \exp \left( - 2 i \tilde \beta \right).
\label{irtys}
\ee
In the Standard Model (SM),
$\tilde \beta = \beta$ and the sine of $2 \beta$ is measured
\cite{browder}
through $CP$ violation in $B^0_d$/$\bar B^0_d \to \psi K_S$:
\be
\sin{2 \beta} = 0.736 \pm 0.049.
\label{iurwq}
\ee
In the SM $\beta$ must be smaller than $\pi / 4$,
hence $\cos{2 \beta}$ is assumed positive.

Together with eq.~(\ref{irtys}),
I shall assume that,
as in the SM,
\be
\frac{\bar A}{A} = \frac{e^{- i \gamma} + z}{e^{i \gamma} + z},
\label{mfjqs}
\ee
where $\gamma$ is another $CP$-violating phase,
which we would like to measure too.
In the SM,
$0 \le \gamma \le \pi - \beta$.
The parameter $z$ represents the `penguin pollution',
an annoying contribution from penguin diagrams
which we must somehow circumvent if we want to get at $\gamma$.

Buchalla and Safir (BS) \cite{safir}
have found a solution to the following problem.
Suppose that
\begin{itemize}
\item one has measured $\sin{2 \tilde \beta}$ and $S$,
\item one has found that $S > - \sin{2 \tilde \beta}$,
\item one assumes the validity of the SM, and
\item one assumes that ${\rm Re}\, z > 0$.
\end{itemize}
Is it then possible to find a lower bound on $\gamma$
stronger than $\gamma \ge 0$?
The solution to this problem,
as given by BS,
is
\be
\gamma > \frac{\pi}{2} - \arctan{\frac
{S - \tau + \tau \sqrt{1 - S^2}}{\tau S + 1 - \sqrt{1 - S^2}}},
\label{BS1}
\ee
where
\be
\tau \equiv
\frac{\sin{2 \tilde \beta}}{1 - \sqrt{1 - \sin^2{2 \tilde \beta}}}
\label{BS2}
\ee
and the square roots in eqs.~(\ref{BS1}) and~(\ref{BS2}) are,
by definition,
positive.

In this Letter I provide a simple derivation of the BS bound,
which does not rely on any assumptions about the quark mixing matrix.
I also consider the realistic situation
where both $S$ and $C$ have been measured;
this allows one to put a stronger bound on $\gamma$
than when one knows only $S$.
Inspired by the result,
quoted by BS,
of a computation of $z$ yielding
\be
\arg{z} = 0.15 \pm 0.25,
\label{mfuit}
\ee
I furthermore consider the situation where one assumes an upper bound
on $\left| \arg{z} \right|$.
Finally,
I apply the BS bound,
and its extensions,
to the most recent measurements of $S$ and $C$ made public
by the experimental collaboration Belle \cite{belle}.

\section{The Buchalla--Safir bound}

I define
\ba
x &\equiv& \lambda \exp \left(2 i \tilde \beta \right) \no
&=& \frac{e^{- i \gamma} + z}{e^{i \gamma} + z}.
\label{x}
\ea
Then,
\be
C = \frac{1 - \left| x \right|^2}{1 + \left| x \right|^2},
\ee
and I furthermore define
\ba
I &\equiv& \frac{2 {\rm Im}\, x}{1 + \left| x \right|^2}, \\
F &\equiv& \frac{\left| 1 - x \right|^2}{1 + \left| x \right|^2} \no
&=& 1 - \frac{2 {\rm Re}\, x}{1 + \left| x \right|^2}.
\ea
Clearly,
\be
0 \le F \le 2
\label{urias}
\ee
and
\be
C^2 + I^2 + F^2 = 2 F.
\ee
Solving eq.~(\ref{x}) for $z$,
one finds
\be
z = - \cos{\gamma} + \frac{- I + i C}{F}\, \sin{\gamma}.
\label{kghqz}
\ee
Equation~(\ref{kghqz}) has an indeterminacy
at the singular point $C = I = F = 0\, \Leftrightarrow\, x = 1$,
\textit{i.e.}$\!$ when $\sin{\gamma} = 0$,
for arbitrary $z$.

From eq.~(\ref{kghqz}) it follows in particular that
\be
F \left( \cos{\gamma} + {\rm Re}\, z \right) + I \sin{\gamma} = 0.
\label{hftwl}
\ee
Equation~(\ref{hftwl}) has been first written down
by Botella and Silva \cite{silva}.
It leads to the bound
\be
\left| {\rm Re}\, z \right| \le \frac {\sqrt{F^2 + I^2}}{F},
\ee
where $\sqrt{F^2 + I^2}$ \emph{is positive by definition}.
The solution to eq.~(\ref{hftwl}) may be written in the form
\be
\gamma = \xi + \chi,
\ee
where (by definition)
\begin{itemize}
\item $\xi$ is independent of ${\rm Re}\, z$, and
\item $\chi = 0$ or $\chi = \pi$ when ${\rm Re}\, z = 0$.
\end{itemize}
One finds
\ba
\cos{\xi} &=& \frac{- I}{\sqrt{F^2 + I^2}},
\label{kghwm}
\\
\sin{\xi} &=& \frac{F}{\sqrt{F^2 + I^2}},
\label{jgyxc}
\ea
and
\ba
\sin{\chi} = \frac{F {\rm Re}\, z}{\sqrt{F^2 + I^2}}.
\label{knvgb}
\ea
While $\xi$ is perfectly defined
by eqs.~(\ref{kghwm}) and (\ref{jgyxc}),
$\chi$ as given by eq.~(\ref{knvgb})
suffers from the twofold ambiguity
\be
\chi \to \pi - \chi.
\label{sfdqr}
\ee
Assuming,
as Buchalla and Safir have done,
that ${\rm Re}\, z > 0$,
we see from eqs.~(\ref{jgyxc}) and~(\ref{knvgb}) that
both $\xi$ and $\chi$ are angles
either of the first or of the second quadrant.
The Buchalla--Safir condition ${\rm Re}\, z > 0$
implies the lower bound on $\gamma$
\ba
\gamma &>& \xi \no
&=& \arccos{\frac{- I}{\sqrt{F^2 + I^2}}},
\label{hfyqo}
\ea
together with $\gamma < \xi + \pi$ too.
Notice that
\be
d \xi = \frac{F d I - I d F}{F^2 + I^2}.
\label{nfopa}
\ee

Equation~(\ref{hfyqo}) does provide a lower bound on $\gamma$ but,
unfortunately,
one has to deal with discrete ambiguities.
These occur because we are able to measure $C$
but unable to measure $I$ and $F$;
rather,
we only know $\sin{2 \tilde \beta}$ and $S$.
Now,
\ba
I &=& \frac{2 {\rm Re}\, \lambda}{1 + \left| \lambda \right|^2}\,
\sin{2 \tilde \beta} + S \cos{2 \tilde \beta},
\\
F &=& 1 - \frac{2 {\rm Re}\, \lambda}{1 + \left| \lambda \right|^2}\,
\cos{2 \tilde \beta} + S \sin{2 \tilde \beta}.
\ea
Assuming that $\sin{2 \tilde \beta}$,
$S$,
and $C$ are known,
there is a fourfold ambiguity in $I$ and $F$,
since the signs of
\ba
\frac{2 {\rm Re}\, \lambda}{1 + \left| \lambda \right|^2}
&=& \sqrt{1 - C^2 - S^2}, \\
\cos{2 \tilde \beta} &=& \sqrt{1 - \sin^2{2 \tilde \beta}}
\label{nsowq}
\ea
remain unknown.
Using eqs.~(\ref{nfopa})--(\ref{nsowq}),
\be
\frac{d \xi}{d C^2} = \frac{- S - \sin{2 \tilde \beta}}
{2 \left( F^2 + I^2 \right) \sqrt{1 - C^2 - S^2}}.
\label{udgfe}
\ee
(Remember that the sign of $\sqrt{1 - C^2 - S^2}$ is,
for the moment,
arbitrary.)

Thus,
given $C$,
$S$,
and $\sin{2 \tilde \beta}$,
there are in reality four different angles $\xi$:
\begin{itemize}
\item $\xi_1$, in which both $\sqrt{1 - C^2 - S^2}$
and $\cos{2 \tilde \beta}$ are positive,
\item $\xi_2$, in which $\cos{2 \tilde \beta}$ is positive
but $\sqrt{1 - C^2 - S^2}$ is negative,
\item $\xi_3$, in which both $\sqrt{1 - C^2 - S^2}$
and $\cos{2 \tilde \beta}$ are negative, and
\item $\xi_4$, in which $\sqrt{1 - C^2 - S^2}$ is positive
but $\cos{2 \tilde \beta}$ is negative.
\end{itemize}
Since $F$ remains invariant,
and $I$ changes sign,
when $\sqrt{1 - C^2 - S^2}$ and $\cos{2 \tilde \beta}$
change sign simultaneously,
we find that $\xi_3 = \pi - \xi_1$ and $\xi_4 = \pi - \xi_2$.
From the assumption that ${\rm Re}\, z > 0$,
and taking into account the indeterminacy in the signs
of $\sqrt{1 - C^2 - S^2}$ and $\cos{2 \tilde \beta}$,
one can only deduce that $\gamma$ must lie
in between $\xi_k$ and $\xi_k + \pi$ for all $k = 1, 2, 3,$ and $4$.

Let us now assume,
with BS,
the validity of the SM.
Then $\cos{2 \tilde \beta}$ is positive
and only the values $\xi_1$ and $\xi_2$
are allowed for $\xi$.
This produces the lower bound
\be
\gamma > {\rm min} \left( \xi_1, \xi_2 \right).
\label{urmso}
\ee
This lower bound is valid in the SM when $C$,
$S$,
and $\sin{2 \tilde \beta}$ are known.
It still depends on $C^2$,
since $\xi_1$ and $\xi_2$ contain $\sqrt{1 - C^2 - S^2}$.
Consideration of eq.~(\ref{udgfe}),
however,
shows that,
when $S > - \sin{2 \tilde \beta}$,
$\xi_1$ decreases and $\xi_2$ increases with increasing $C^2$.
Moreover,
at the maximum allowed value of $C^2$,
\textit{i.e.}$\!$ when $C^2 = 1 - S^2$,
one has $\xi_1 = \xi_2$,
since in general $\xi_1$ and $\xi_2$ only differ
through the sign of $\sqrt{1 - C^2 - S^2}$,
and that square root becomes zero when $C^2 = 1 - S^2$.
This immeadiately leads to the BS bound:
if $S > - \sin{2 \tilde \beta}$,
then $\gamma > \xi_2 \left( C^2 = 0 \right)$.
It can be shown \cite{silva} that,
though different in appearance,
this bound coincides with the one in eq.~(\ref{BS1}).

One thus concludes that,
if one assumes that $\cos{2 \tilde \beta} > 0$,
then
\be
\left\{
\begin{array}{rcl}
\gamma > \xi_2 \left( C^2 = 0 \right) &\Leftarrow&
S > - \sin{2 \tilde \beta}, \\
\gamma > \xi_1 \left( C^2 = 0 \right) &\Leftarrow&
S < - \sin{2 \tilde \beta}.
\end{array}
\right.
\ee
This may be put in a more transparent way if one defines
\ba
\varphi &\equiv& \frac{1}{2}\, \arcsin S, \label{jfgew} \\
\alpha &\equiv& \pi - \tilde \beta - \gamma.
\ea
The lower bound on $\gamma$ may then be rewritten
as an upper bound on $\alpha$:
\be
\left\{
\begin{array}{lcl}
\alpha < {\displaystyle \frac{\pi}{2}} - \varphi &\Leftarrow&
\varphi > - \tilde \beta, \\*[2mm]
\alpha < \pi + \varphi &\Leftarrow& \varphi < - \tilde \beta.
\end{array}
\right.
\ee
The discontinuity of the bound at $\varphi = - \tilde \beta$
should not come as a surprise.
The point $C = 0$,
$S = - \sin{2 \tilde \beta}$ allows the singularity
$C = I = F = 0$ referred to earlier.
When $C = I = F = 0$,
$\gamma$ may be either $0$ or $\pi$,
independently of any assumption on $z$.
Therefore no lower bound on $\gamma$ may be derived
if the experimentally allowed region for $C$ and $S$
includes that point.

It should be stressed that this derivation
of the Buchalla--Safir bound on $\gamma$,
or on $\alpha$,
contains basically no physical assumptions.
Only eqs.~(\ref{iurty})--(\ref{irtys}) and~(\ref{mfjqs}),
together with $\cos{2 \tilde \beta} > 0$ and ${\rm Re}\, z > 0$,
are assumed.
No assumptions are needed about the physics
contained in the decay amplitudes,
about the quark mixing matrix,
or,
indeed,
about anything else;
the sole crucial assumption is ${\rm Re}\, z > 0$.
The Buchalla--Safir bound is purely algebraic.

I now return to the general case where one does not assume the SM.
Then,
$\gamma$ may be either positive or negative and,
from the assumption that ${\rm Re}\, z > 0$,
it is only possible to produce a lower bound on $\left| \gamma \right|$,
never on $\gamma$ itself.
Indeed,
given the fourfold ambiguity in the determination of $F$ and $I$,
and the twofold ambiguity in the determination of $\chi$---see
eq.~(\ref{sfdqr})---there are eight solutions
to eq.~(\ref{hftwl}) for $\gamma$.
Since,
when $\sqrt{1 - C^2 - S^2}$ and $\cos{2 \tilde \beta}$
change sign simultaneously,
$I$ changes sign while $F$ does not change,
it is obvious from eq.~(\ref{hftwl}) that those eight solutions
pair in four sets through the transformation $\gamma \to - \gamma$.
Therefore,
only a bound on $\left| \gamma \right|$ is possible.
Now,
computing
\be
\tan^2{\xi_1 \left( C^2 = 0 \right)}
- \tan^2{\xi_2 \left( C^2 = 0 \right)}
= \frac{- 4 \sqrt{1 - S^2} \sqrt{1 - \sin^2{2 \tilde \beta}}}
{\left( \sin{2 \tilde \beta} - S \right)^2},
\ee
where the square roots in the right-hand side are positive by definition,
one finds that $\left| \tan{\xi_1 \left( C^2 = 0 \right)} \right|$
is always smaller than $\left| \tan{\xi_2 \left( C^2 = 0 \right)} \right|$.
Hence,
\be
\left| \gamma \right|
> \arctan{\left| \tan{\xi_1 \left( C^2 = 0 \right)} \right|}.
\ee
Using again $\varphi$ as defined in eq.~(\ref{jfgew}),
one concludes that
\be
\left| \gamma \right| > \left| \tilde \beta + \varphi \right|,
\ee
which is valid in any model provided ${\rm Re}\, z > 0$---and
provided the basic equations~(\ref{iurty})--(\ref{irtys})
and~(\ref{mfjqs}) hold,
of course.

\section{Assuming an upper limit on $\left| \arg{z} \right|$}

In their work \cite{safir},
Buchalla and Safir have quoted the result of a computation
(in the context of the Standard Model)
of $z$ as yielding the result in eq.~(\ref{mfuit}).
They have thereby justified their assumption ${\rm Re}\, z > 0$.
In this section I shall consider a different assumption,
\be
\left| \cot{\arg{z}} \right| > L,
\label{raita}
\ee
where $L$ is some positive number.
Clearly,
this assumption is complementary to ${\rm Re}\, z > 0$;
while ${\rm Re}\, z > 0$,
by itself alone,
leaves $\cot{\arg{z}}$ completely arbitrary,
eq.~(\ref{raita}),
by itself alone,
does not provide any information on whether ${\rm Re}\, z$
is positive or negative.
If $L$ is,
for instance,
taken equal to $1$,
then eq.~(\ref{raita}) is well justified by eq.~(\ref{mfuit}).

In order to find the consequences of the assumption in eq.~(\ref{raita}),
I return to eq.~(\ref{kghqz}) and therefrom derive that
\be
C \cot{\arg{z}} + F \cot{\gamma} + I = 0.
\ee
Hence,
\be
\left| \cot{\arg{z}} \right| > L
\quad \Leftrightarrow \quad
\cot{\gamma} < \frac{- I - L \left| C \right|}{F}
\quad {\rm or} \quad
\cot{\gamma} > \frac{- I + L \left| C \right|}{F}.
\ee
Clearly,
this condition makes smaller the range for $\gamma$
allowed by ${\rm Re}\, z > 0$ alone;
that range,
remember,
is given by $\xi < \gamma < \xi + \pi$,
where $\xi$ belongs either to the first or to the second quadrant
and $\cot{\xi} = - I / F$.

Let us now assume the validity of the SM.
Then $\gamma \le \pi - \beta$ and the relevant bound on $\gamma$
following from eq.~(\ref{raita}) is the lower bound
\ba
\cot{\gamma} &<& \frac{- I - L \left| C \right|}{F} \no
&=& \frac{- \sqrt{1 - C^2 - S^2} \sin{2 \tilde \beta}
- S \cos{2 \tilde \beta} - L \left| C \right|}
{1 - \sqrt{1 - C^2 - S^2} \cos{2 \tilde \beta}
+ S \sin{2 \tilde \beta}}.
\label{urtyc}
\ea
This bound depends on the measured values of $C$,
$S$,
$\sin{2 \tilde \beta}$ and,
besides,
since $\cos{2 \tilde \beta}$ is positive in the SM,
it depends on the sign of $\sqrt{1 - C^2 - S^2}$.

\section{Application to the Belle results}

\begin{figure}[ht]
\centering
\includegraphics[clip,height=90mm]{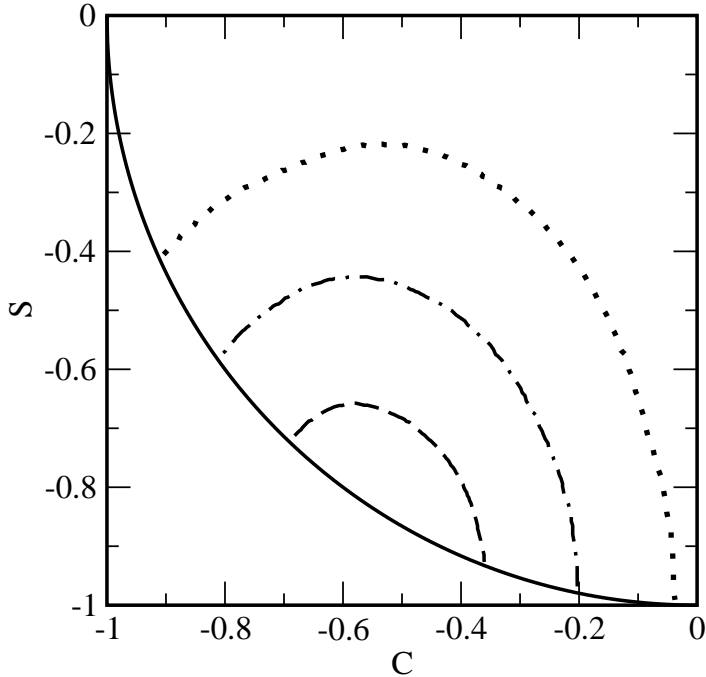}
\caption{The latest results of the Belle Collaboration for $S$ and $C$.
The full line bounds the circle defined by the condition $C^2 + S^2 \le 1$.
Within that circle,
the dashed line bounds the region allowed by Belle at 68.3\% C.L., 
the dot-dashed line bounds the region allowed at 95.45\% C.L., 
and the dotted line bounds the region allowed at 99.73\% C.L.}
\label{fig1}
\end{figure}

The BS bound applies to the situation where $S$ has been measured
while $C$ remains unknown but,
in reality,
both the BABAR and Belle Collaborations are able to measure $S$ and $C$
simultaneously and with comparable accuracy.
Early results made public by BABAR \cite{babar} are
\be
\begin{array}{rcl}
S &\in& \left[ -0.54,\ 0.58 \right], \\
C &\in& \left[ -0.72,\ 0.12 \right]
\end{array}
\ee
at 90\% Confidence Level (C.L.).
In this section I shall rather use the latest results
by the Belle Collaboration \cite{belle}.
Belle measures $S$ and $C$ to be both negative
and not satisfying the constraint $S^2 + C^2 \le 1$;
enforcing the latest constraint,
the Belle Collaboration has presented the allowed regions for $C$ and $S$
displayed in fig.~\ref{fig1}.
The point $C = 0$,
$S = - \sin{2 \beta}$ is disallowed at 99.9157 C.L.,
and therefore setting a BS lower bound on $\gamma$ is possible.
Assuming the SM,
the lower bound on $\gamma$ that I shall consider
is given by the inequality~(\ref{urtyc}),
where $\sqrt{1 - C^2 - S^2}$ may be either positive or negative---we
must use,
for each pair of values for $S$ and $C$,
the sign of $\sqrt{1 - C^2 - S^2}$ yielding the less stringent bound.
I shall assume fixed values for $\sin{2 \tilde \beta} = 0.736$
and $\cos{2 \tilde \beta} = + \sqrt{1 - 0.736^2}$.
For $L$ I shall take the four values
$L=0$---the case relevant for the BS bound,
where ${\rm Re}\, z > 0$,
but no lower bound on $\left| \cot{\arg{z}} \right|$,
is assumed---and $L = \cot{0.9}$,
$\cot{0.65}$,
and $\cot{0.4}$,
corresponding to the $3\sigma$,
$2\sigma$,
and $1\sigma$ bounds,
respectively,
following from eq.~(\ref{mfuit}).

I performed scans of the allowed regions
in the $\left( C, S \right)$ plane
advocated by the Belle Collaboration.
For each value of the pair $\left( C, S \right)$,
and for each value of $L$,
I computed the corresponding lower bound on $\gamma$.
The results are the following.
If one takes the 68.3\% C.L.\ domain of Belle,
then $\gamma > 21.8^\circ$ if $L = 0$,
$\gamma > 42.3^\circ$ if $L = \cot{0.9}$,
$\gamma > 58.3^\circ$ if $L = \cot{0.65}$,
and $\gamma > 93.6^\circ$ if $L = \cot{0.4}$.
When one uses the the region allowed by Belle at 95.45\% C.L.,
one obtains $\gamma > 12.3^\circ$ if $L = 0$,
$\gamma > 24.1^\circ$ if $L = \cot{0.9}$,
$\gamma > 33.9^\circ$ if $L = \cot{0.65}$,
and $\gamma > 53.7^\circ$ if $L = \cot{0.4}$.
Considering at last the 99.73\% C.L.\ limits of Belle,
one gets $\gamma > 3.6^\circ$ if $L = 0$,
$\gamma > 6.6^\circ$ if $L = \cot{0.9}$,
$\gamma > 8.9^\circ$ if $L = \cot{0.65}$,
and $\gamma > 12.5^\circ$ if $L = \cot{0.4}$;
these very loose bounds reflect the proximity to this region
of the point $C = 0$,
$S = - \sin{2 \beta}$,
for which no lower bound on $\gamma$ is possible any more.

It is evident from the results above that
assuming $\left| \cot{\arg{z}} \right| > L$,
with a non-zero $L$,
may greatly improve the lower bound on $\gamma$ that
one obtains from the BS condition ${\rm Re}\, z > 0$ alone.

\section{Conclusions}

I have shown in this Letter that the Buchalla--Safir
lower bound on $\gamma$
is a purely algebraic consequence of the assumption ${\rm Re}\, z > 0$;
the latter assumption follows from a computation of $z$
within the Standard Model but,
after that computation,
the derivation of the BS bound itself requires no physics.
I have emphasized that a better lower bound on $\gamma$
may be obtained if one considers that,
besides $S$,
also $C$ is known.
I have improved the BS bound by assuming,
above and beyond ${\rm Re}\, z > 0$,
a lower bound on $\left| \cot{\arg{z}} \right|$.
I have emphasized the fact that the presence,
within the experimentally allowed region,
of the point $\left(S, C \right) = \left( - \sin{2 \beta}, 0 \right)$,
prevents one from putting a lower bound on $\gamma$.
I have applied the derived bounds
to the $\left( S, C \right)$ domains allowed
by the most recent results made public by the Belle Collaboration.

\vspace*{5mm}

\paragraph{Acknowledgements} I am grateful to Jo\~ao Paulo Silva
for introducing the BS bound to me,
and for discussions;
his paper with Botella \cite{silva}
motivated and strongly influenced the present Letter.
I am also grateful to Jorge Crispim Rom\~ao,
who has helped me willingly for several times in software matters.
This work has been supported by the Portuguese
\textit{Funda\c c\~ao para a Ci\^encia e a Tecnologia}
under the project CFIF--Plurianual.

\end{document}